\documentclass[aps,prd,nofootinbib,showpacs,preprintnumbers,longbibliography,amssymb,11pt]{revtex4-1} 
\usepackage{url,subcaption,comment}
\usepackage{hyperref}

\usepackage{graphicx}
\usepackage{epsfig}
\usepackage{dcolumn}
\usepackage{bm}
\usepackage{amsmath}
\usepackage{amssymb}

\graphicspath{{./Figures/}}

\newcommand{\eqnref}[1]{Eq.~(\ref{eqn:#1})}

\newcommand{\figref}[1]{Fig.~\ref{fig:#1}}

\usepackage[compat=1.1.0]{tikz-feynman}
\tikzset{rubout/.style={preaction={draw=white,line width=3pt}}}

\begin{document}

\baselineskip=18pt \pagestyle{plain} \setcounter{page}{1}

\vspace*{-1cm}

\preprint{MITP-23-038}

\begin{center}

\title{Primer on Axion Physics}


\author{Felix Yu}\email{yu001@uni-mainz.de}
\affiliation{\footnotesize \sl PRISMA$^+$ Cluster of Excellence \& Mainz Institute for Theoretical Physics\\
Johannes Gutenberg University, 55099 Mainz, Germany}


\begin{abstract}
I review the canonical axion potential, with an emphasis on the field
theory underlying radial and angular modes of complex scalar fields.
I present the explicit calculation of the instanton-induced breaking
of the Goldstone field direction necessary to derive the canonical
axion mass and decay constant relation.  
The primer is
intended to serve an audience with elementary quantum field theory
expertise.

\end{abstract}

\maketitle
\end{center}



{\small
\hypersetup{linktocpage} 
\tableofcontents
\hypersetup{linkcolor=red} 
}







\section{Introduction}
\label{sec:Introduction}

Axions are hypothesized particles originally proposed to resolve the
strong CP problem of the Standard Model.  The essential mechanism of
the axion solution is that the quantum chromodynamics (QCD)
instanton-generated potential for the axion is minimized when the
vacuum expectation value of the axion exactly cancels the (unknown)
original $\bar{\Theta}$ parameter, leaving an effective
$\bar{\Theta}_{\text{eff}}$ parameter that is vanishing.  In this
review, we will see the fundamental distinction between the axion
potential and other, more common scalar potentials usually considered in
quantum field theories, such as the Higgs potential of the Standard
Model.  We will also analyze the field theory from the modern
viewpoint of Peccei-Quinn (PQ) symmetry, again making a distinction
between the role of PQ symmetries relevant for axion physics and the
more traditional electroweak gauge symmetry emphasized in Higgs
physics.  The goals of this primer are to derive and motivate the
axion potential as a characteristic and benchmark model of
pseudo-Nambu-Goldstone boson (pNGB) Lagrangians as well as to develop
the basic phenomenological signals of axion and axion-like particles
as ultralight dark matter candidates.  Several excellent reviews of axion physics include~\cite{Kim:2008hd, Kawasaki:2013ae, Graham:2015ouw, Irastorza:2018dyq, Hook:2018dlk, DiLuzio:2020wdo, Reece:2023czb}, as well as the ever-evolving Particle Data Group review~\cite{ParticleDataGroup:2022pth}, which expertly cover the breadth of axion physics in cosmology, particle physics, and field theory at a high level.  In contrast, this primer is designed to build axion physics from the ground up, assuming only a beginning level of quantum field theory knowledge.

The mathematical framework for fundamental laws of Nature, in
particular the Standard Model, is quantum field theory, which is based
on the quantum action $S = \int d^4 x \mathcal{L}$, where the
Lagrangian $\mathcal{L}$ (technically, the Lagrangian density, but we
always call it the Lagrangian) is required to respect Poincar\'{e}
invariance.  Poincar\'{e} symmetry is Lorentz symmetry $\oplus$
translation symmetry, and is required to establish the familiar
energy-momentum conservation laws as well as the characterization of
particles according to half-integer representations for spin-angular
momentum.  This motivates starting with the simplest quantum field theory, namely that of a scalar field whose excitations are scalar particles with spin $0$.

\section{Scalar field theory}
\label{sec:scalars}

We begin with the field theory of a single scalar field $\phi$.
Scalar fields assign a pure real number to every spacetime point, and
thus they take the spacetime coordinate as their argument.  For a
scalar field with mass $m$ and no interactions, we can write the
Lagrangian (density) as
\begin{align}
\mathcal{L} = \frac{1}{2} \left( \partial_{\mu} \phi \right)^2 -
\frac{1}{2} m^2 \phi^2 \ .
\label{eqn:realphi}
\end{align}
Analyzing the Euler-Lagrange equation of motion dictated by this
Lagrangian will recover the familiar Klein-Gordon equation governing
free scalars with a mass $m$, assuming $m > 0$.

If there is a second scalar field with exactly the same mass, then the
two fields can be combined into a single {\it complex} scalar field
$\Phi$, where the real part of $\Phi$ is the number assigned by the
original $\phi$ field and the imaginary part of $\Phi$ represents the
second scalar field.  Clearly, the Lagrangian in this case can be
written as
\begin{align}
\mathcal{L} = \left| \partial_{\mu} \Phi \right|^2 - m^2 | \Phi|^2 \ .
\end{align}
Note the prefactors of the kinetic and mass term are twice as large as
the real scalar Lagrangian, reflecting the existence of two distinct
real degrees of freedom.

To connect to Higgs physics and the phenomenon of spontaneous symmetry
breaking, we can now add a quartic interaction.  We have
\begin{align}
\mathcal{L} = \left| \partial_{\mu} \Phi \right|^2 - m^2 | \Phi|^2 \ -
\lambda |\Phi|^4 \ .
\end{align}
We remark that the Lagrangian is always the kinetic energy minus potential
energy, and so we can analyze the potential
\begin{align}
V = m^2 | \Phi|^2 + \lambda | \Phi|^4 \ .
\end{align}
In order for the potential to be bounded from below, we require
$\lambda > 0$, but we now have two possibilities for the sign of
$m^2$.  If $m^2 > 0$, the origin in field space is stable and the
field has no vacuum expectation value (vev).  If $m^2 < 0$, the origin
in field space is unstable and the equation of motion for the scalar
is extremized by a nonzero vev, $|\Phi|^2 = v^2 = -m^2 / \lambda$.
In either case, as we go to large excursions in field space, the
potential energy grows because of $\lambda > 0$, but the behavior near
the origin in field space will distinguish the two potentials.  As an
aside, if we coupled the complex scalar field to a $U(1)$ gauge
symmetry, replacing $\partial_{\mu}$ by a covariant derivative
$D_{\mu}$ in the equation above, then the above Lagrangian would be
the scalar part of the Abelian Higgs model, and the vev of the Higgs
field would give a mass for the $U(1)$ gauge boson.

Importantly, the vacuum state of the theory is controlled by one
constraint equation, $| \Phi|^2 = v^2$, and therefore we have one
degree of freedom that is {\it unconstrained} by the vacuum
requirement.  This is most evident when we change from a ``Cartesian''
field space parametrization to a ``polar'' parametrization,
\begin{align}
\Phi = \left( \frac{v + h}{\sqrt{2}} \right) \exp (i a / v) \ ,
\label{eqn:phiPolar}
\end{align}
where $h$ and $a$ are the {\it radial} and {\it angular} modes of the
complex field $\Phi$.  When we write the original Lagrangian with this
parametrization, we get
\begin{align}
\mathcal{L} = \frac{1}{2} \left( \partial_\mu h \right)^2 +
\frac{1}{2} \left( \partial_\mu a \right)^2 + \frac{1}{4} \lambda v^4
- \lambda v^2 h^2 - \lambda v h^3 - \frac{1}{4} \lambda h^4 \ ,
\label{eqn:expandedHiggs}
\end{align}
where we notice that the angular field $a$ does not appear in the
potential anymore while the radial field $h$ has a mass term $m_h =
\sqrt{2 \lambda} v$ as well as cubic and quartic self-interactions.
We also remark the appearance of a constant energy (density) term in
the potential, which plays no role in the dynamics at hand but would
be relevant for the cosmological constant problem.  The key point is
that the angular field is a completely free and massless
(pseduo)scalar field, and its (currently vanishing) dynamics are
independent of the dynamics of the radial mode.

\begin{figure}
\includegraphics[width=0.8\textwidth]{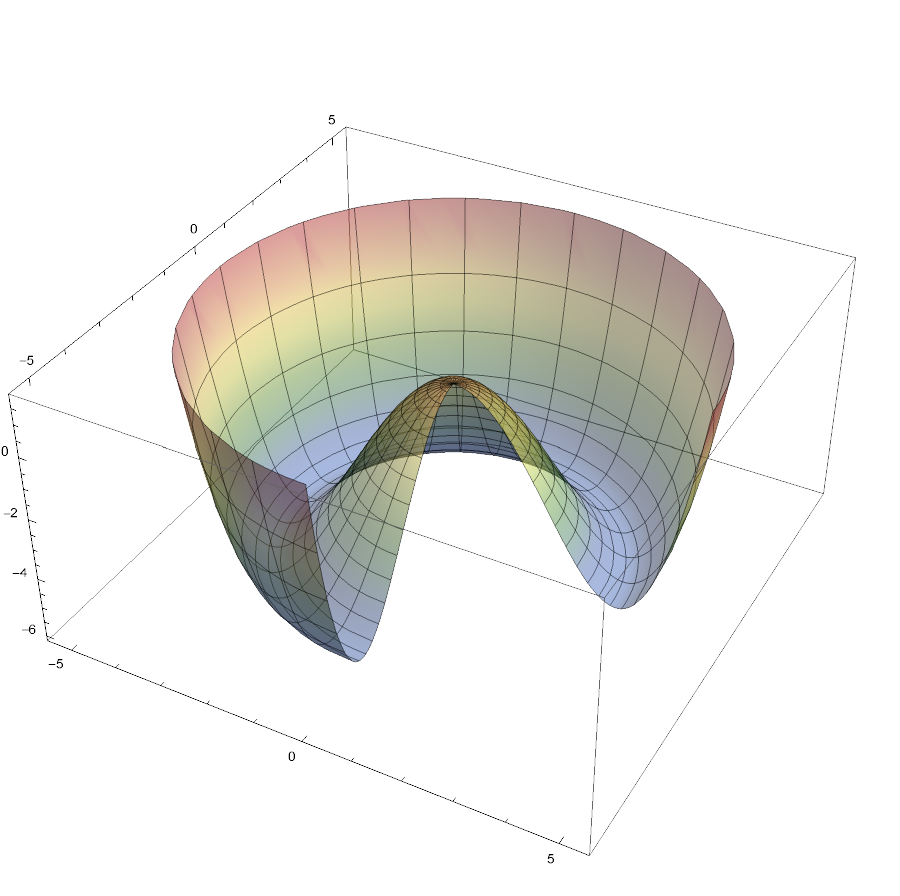}
\caption{Schematic of the typical Higgs potential from a negative
  mass-squared term and a positive quartic term.  The rotational
  symmetry is manifest for every point in field space and represents
  the Goldstone field direction.}
  \label{fig:Higgspot}
\end{figure}

We remark that the pseudoscalar nature of the $a$ field in~\eqnref{expandedHiggs} is only meaningful in the context of the embedding as the angular mode of the original complex scalar field $\Phi$.  In particular, comparing~\eqnref{realphi} and the $a$-dependent term in~\eqnref{expandedHiggs} shows that~\eqnref{realphi} reduces to the $a$-dependent kinetic term if we neglect the mass $m$ in~\eqnref{realphi} and relabel $\phi$ to $a$.  While this relabeling seems entirely safe given that the Lagrangians now coincide, we have to recall that $a$ is defined as a {\it compact} field direction via~\eqnref{phiPolar} while the original free scalar field $\phi$ lies on a {\it noncompact} field space.  Formally, this distinction about compactness is reflected in the fact that the action $S$ has an extra periodic redundancy for $a \to a + 2 \pi v$  but not for a massless $\phi$, as evident from the polar decomposition in~\eqnref{phiPolar}.  

Importantly, the ``zero point" in field space for $a$ is entirely degenerate and lies along the entirety of the circle in~\figref{Higgspot}, as a result of the vev only being defined by its square.  This is self-consistent with the Goldstone nature of $a$ having a continuous shift symmetry in~\eqnref{expandedHiggs}, where any finite shift of $a$ leaves the Lagrangian invariant.  In fact, the absence of an origin in field space is crucial for the axion solution to the strong CP problem, since it allows the axion to render any arbitrary $\bar{\Theta}$ starting value to (effectively) zero via the PQ mechanism.

\section{Instantons in Yang-Mills theory and the $\Theta$ vacuum energy}
\label{sec:instantons}

In this section, we review the non-perturbative calculation of the
instanton-induced vacuum energy of Yang-Mills theory.  This
calculation is taken from the tour-de-force publicaation of 't
Hooft~\cite{tHooft:1976snw} and related literature on
instantons~\cite{Callan:1976je, Callan:1977gz}.

At the close of the last section, we demonstrated that the angular
mode in a Higgsed potential is a free field with only a kinetic term.
In more formal terms, the angular mode is a pseudoscalar field that
has a continuous shift symmetry, meaning that any shift in the field has no effect on the Lagrangian.
In particular, the field has no mass terms and no interactions, since
masses and interactions violate the continuous shift symmetry.  We remark that the generic nature of Goldstone bosons arising in ultraviolet extensions of the Standard Model can typically be mapped to the field theory at hand, although the key requirements to satisfy are that the ultraviolet fields exhibit Goldstones traversing a compact field space and the ultraviolet continuous symmetries are not gauged.\footnote{Goldstones arising from noncompact continuous symmetries are called dilatons.}  This will be relevant for the extension of the axion parameter space to axion-like particles.

The continuous shift symmetry of the angular mode is
explicitly broken if the symmetry has an anomaly.  Anomalies are, by
definition, quantum origins of symmetry breaking, and in the context
of the axion, originate from the fact that the continuous shift
symmetry for the pseudoscalar field is broken to a periodic shift
symmetry from its axial coupling to fermions carrying color charges.  The fact that the continuous shift symmetry has a nonzero anomaly with respect to the SM color gauge group is the defining property of Peccei-Quinn symmetries, and this $U(1)_{\text{PQ}} \times SU(3)_c^2$ anomaly explicitly breaks $U(1)_{\text{PQ}}$ to a periodic shift symmetry and reflected in the coupling of the axion field to the dual field strength tensor of QCD.  At low energies, the dual field strength tensor of
QCD leads to an instanton-induced potential for the axion, which we
will derive in this section.  We remark that the periodicity condition~\eqnref{phiPolar} and the color anomaly quantizes the vacuum structure of the axion field space according to a topological winding number, permitting domain wall solutions of the $a$ field interpolating between energetically degenerate but topologically distinct vacuum states.

\subsection{Yang-Mills theory and Instantons}
\label{subsec:instantons}
Instantons are a feature of Yang-Mills theories.  The Yang-Mills
action for an $SU(N)$ gauge theory in Minkowski metric is
\begin{align}
\mathcal{L} = -\frac{1}{4} G_{\mu \nu}^a G^{\mu \nu, a} ,
\label{eqn:YangMills}
\end{align}
where $G_{\mu \nu}^a \equiv \partial_\mu A_\nu^a - \partial_\nu
A_\mu^a + g_s f^{abc} A_\mu^b A_\nu^c$.
The simplest method for deriving the instanton solution follows
Belavin, Polyakov, Schwartz, and Tyupkin (BPST)~\cite{Belavin:1975fg}.  

First, we define $\tilde{G}^a = \frac{1}{2} \epsilon_{\mu \nu \rho \sigma} G^a_{\rho \sigma}$ for the totally antisymmetric tensor $\epsilon_{\mu \nu \rho \sigma}$ and $\epsilon_{0123} = 1$.  Then, since $G \tilde{G}$ is a total derivative, and since we are looking for classical field solutions to the action with minimized action, a unit instanton satisfies the quantization constraint
\begin{align}
\int d^4 x_E \frac{g_s^2}{32 \pi^2} G \tilde{G} = 
\int d^4 x_E \frac{g_s^2}{64 \pi^2} \epsilon^{\mu \nu \rho \sigma} 
G_{\mu \nu}^a G_{\rho \sigma}^a = 1 \ ,
\label{eqn:GGtildeunit}
\end{align}
where $x_E$ is the Euclidean spacetime coordinate.  We will relate this quantization constraint to the Chern-Simons topological current shortly.

The procedure from
BPST is to consider the minimal action of the gauge fields arising from the inequality
\begin{align}
0 &\leq \int d^4 x_E \left( G_{\mu \nu}^a \pm \frac{1}{2} \epsilon_{\mu
  \nu \rho \sigma} G_{\rho \sigma}^a \right)^2 \nonumber \\
&= \int d^4 x_E \left( G_{\mu \nu}^a \pm \epsilon_{\mu \nu \rho
  \sigma} G_{\mu \nu}^a G_{\rho \sigma}^a + \frac{1}{4} \epsilon_{\mu
  \nu \rho \sigma} \epsilon_{\mu \nu \eta \theta} 
G_{\rho \sigma}^a G_{\eta \theta}^a \right) \nonumber \\
&= \int d^4 x_E \left[ 2 (G_{\mu \nu})^2 \pm 2 \left( \frac{1}{2}
  \epsilon_{\mu \nu \rho \sigma} G_{\mu \nu}^a G_{\rho \sigma}^a
  \right) \right] \ .
\end{align}
In order to saturate the extremization condition, we must solve the first-order differential equation
\begin{align}
G_{\mu \nu}^a = \pm  \frac{1}{2} \epsilon_{\mu \nu \rho \sigma} G_{\rho \sigma}^a \ ,
\label{eqn:selfduality}
\end{align}
for $A_\mu$ field solutions, where $+$ denotes self-dual fields and $-$ indicates anti-self-dual
fields.  The explicit profile of the instanton solution is essentially a
monopole in group space, and can be written using 't Hooft
symbols~\cite{tHooft:1976snw, Shifman:2022shi}.  For example, an $SU(2)$ instanton can be written in temporal gauge ( $A_0 (\vec{x}) = 0$ ) as the gauge field 
\begin{align}
    A_i (\vec{x}) = i U_1(\vec{x}) \partial_i U_1^{\dagger} (\vec{x}) , 
    \label{eqn:Ai_instanton}
\end{align}
where the $i = 1, 2, 3$ subscripts refer to $\hat{x}$, $\hat{y}$, and $\hat{z}$ components of the vector field and $U_1 (\vec{x})$ is the $SU(2)$-valued group matrix 
\begin{align}
U_1(\vec{x}) = \exp \left( i \pi \vec{x} \cdot \vec{\tau} / \sqrt{ |\vec{x}|^2 + \rho^2} \right) , 
\label{eqn:Ugauge_instanton}
\end{align}
where $\vec{\tau} = \vec{\sigma} / 2$ are the $SU(2)$ generators and $\rho$ is a parameter characterizing the instanton size.  This prescription generalizes to higher winding numbers by replacing $U_1$ by $U_n \equiv U_1^n$ for the desired winding number $n$.  An additional complication arises when considering instanton solutions for groups larger than $SU(2)$, but a fortunate simplification arises from the fact that all $SU(N)$ groups share the same homotopy classifcation $\Pi_3 (SU(N)) = \mathbb{Z}$, which essentially means that all instanton solutions of a given winding number in an $SU(N)$ group are smoothly deformable into each other.  Thus, it is sufficient to characterize instanton solutions in an arbitrary $SU(N)$ group by first considering a convenient $SU(2)$ subgroup and then augmenting the solutions by an appropriate ``index of embedding" that keeps track of the dimensionality of the larger group space~\cite{Shifman:2022shi, Csaki:1998vv}.

We remark that a standard introduction to instantons motivates~\eqnref{Ai_instanton} as gauge-equivalent to $0$.  From this perspective, we note that (still in temporal gauge) trivial $A_i = 0$ solutions to the field equations obviously minimize the Yang-Mills action.  Yet since the action is defined by an integral over (Euclidean) spacetime, we should simultaneously consider all gauge field configurations that are equivalent to the trivial $A_i = 0$ solution, which motivates us to consider ``pure gauge" field configurations for $A_i$ as written in~\eqnref{Ai_instanton}.  Imposing the (anti-)self-duality condition from~\eqnref{selfduality} then leads to the BPST (anti-)instanton solution.

\subsection{The $\Theta$-vacuum in Yang-Mills}
\label{subsec:Theta}

The main consequence of instanton solutions in Yang-Mills theory is its impact on the vacuum structure.  We first illustrate this point by noting the unit instanton solution satisfying~\eqnref{GGtildeunit} corresponds to the flow of the Chern-Simons current~\cite{Shifman:2022shi}
\begin{align}
    K^\mu = 2 \epsilon^{\mu \nu \alpha \beta} \left(
    A_\nu^a \partial_\alpha A_\beta^a + \frac{g}{3} f^{abc} A_\nu^a A_\alpha^b A_\beta^c
    \right) \ ,
    \label{eqn:Kcurrent}
\end{align}
where the charge of the $K^\mu$ current is the usual integral of $K^0$ over three-dimensional space,
\begin{align}
    \mathcal{K} = \frac{g^2}{32 \pi^2} \int d^3 x K^0 (x) \ .
    \label{eqn:Kcharge}    
\end{align}
While the Chern-Simons current is not gauge-invariant and hence not a physical current, its divergence is gauge-invariant and actually the familiar $G \tilde{G}$ operator,
\begin{align}
    \partial_\mu K^\mu = G \tilde{G} ,
\label{eqn:CSdiv}
\end{align}
which allows us to identify the unit instanton as carrying Chern-Simons topological charge.  Consequently, the vacuum state of Yang-Mills theory is an infinitely degenerate vacuum state with ``pre-vacua" labeled by Chern-Simons charge, as required from the homotopy classification.

We now calculate the vacuum energy of $SU(2)$ Yang-Mills theory, which can be
obtained using the functional integral method.  We start with
\begin{align}
\exp \ ( {-\mathcal{E} ~ V \cdot T} )&= \int \mathcal{D}A \exp (-S_E \left[A \right] ) \ ,
\end{align}
where $S_E[A]$ is the Euclidean action corresponding to~\eqnref{YangMills}  and $\mathcal{E}$ is the vacuum
energy density and $V \cdot T$ is the volume of Euclidean spacetime.  
The functional integral is a sum over topological
sectors with winding numbers $w$ as 
\begin{align}
\exp \ ( {-\mathcal{E} ~ V \cdot T} ) &= 1 \times \int 
\left[ \mathcal{D}A \right]_{w = 0} \exp \left[
-\int \frac{1}{2} \delta A \cdot (-\partial^2) \delta A \right] 
\nonumber \\
&+
\exp (-8 \pi^2 / g^2) \times 2 \int \left[ \mathcal{D}A \right]_{w = \pm 1} \exp \left[
-\int \frac{1}{2} \delta A \cdot (-\partial^2) \delta A \right] 
\nonumber \\
&+ \ldots \ ,
\end{align}
where the $\ldots$ indicate a sum over higher topological sectors that
are further suppressed by higher powers of the instanton action.  The
remaining path integral is calculated over variations $\delta A$ centered around
the instanton background with fixed $w$~\cite{tHooft:1976snw, Callan:1976je, Callan:1977gz} and essentially distills into an evaluation of the determinantal operator associated with the eigenvalues of $-\partial^2$, which must be regularized and renormalized.  Eschewing the technical aspects of the calculation, the evaluation of the path integral modifies the $1/g^2$ coupling in instanton action prefactor into a running coupling, where $1 / g^2 (\rho)$ is to be evaluated at the appropriate instanton size.  

In addition, the overall path integral suffers divergences from the integration of symmetries of the instanton solution.  Namely, there are eight transformations that leave the classical solution invariant: four  translations, one  dilatation and three $SU(2)$ rotations. The three $SU(2)$ rotations are handled by an integration over Euler angles specifying the orientation of the unit instanton solution in group space.  We parametrize the translations by a shift of $x_E^j \to x_E^j + x_0^j$, $j = 1, 2, 3, 4$, which serve to shift the center of the instanton solution in~\eqnref{Ai_instanton} and~\eqnref{Ugauge_instanton} but do not change its topological index, and we have already introduced the instanton size parameter $\rho$ in~\eqnref{Ugauge_instanton}.  Simply stated, the path integral is a discrete sum over all instantons of a given topological index, and thus the zero mode redundancies of solutions with the same topological index must be parametrized as explicit integration variables.
Hence, the zero modes for translation symmetry and dilatation symmetry are left as indefinite integrals over $d^4 x_0$ and $d\rho$, giving 
\begin{align}
\exp \ ( {-\mathcal{E} ~ V \cdot T} ) &= \text{ const.} \left( 1+ 2 \int
d^4 x_0 \int \frac{d \rho}{\rho^5} ~ C \frac{1}{g^8} \exp 
\left( -8\pi^2 / g^2(\rho) \right) + \ldots \right) \ ,
\label{eqn:d4xdrho}
\end{align}
where $C  =   2^{10}  \pi^6
e^{7.0539\ldots.}$  from  the  classic 't  Hooft  calculation~\cite{tHooft:1976snw}.  The  $\frac{1}{\rho^5}$   comes   from dimensional  analysis, and the $1 / g^8$
factor  comes from  the  Jacobian  for change  of  variables from  the
integral   over  zero  modes.  Subsequent terms corresponding to $n_+$ instantons and $n_-$ anti-instantons, denoted by $\ldots$ in~\eqnref{d4xdrho}, are given by
\begin{align}
\frac{1}{n_+ ! ~ n_- !} \left[
\int d^4 x_0 \int \frac{d \rho}{\rho^5} ~ C \frac{1}{g^8} \exp
(-8\pi^2 / g^2(\rho)) \right]^{n_+ + n_-} \ ,
\label{eqn:dilute}
\end{align}
which is the basis for the dilute gas approximation where the
instantons are well-separated.  We remark that as $\rho$  becomes
large,  $g^2(\rho)$ grows and  the semiclassical  approximation breaks
down.

We can now reshuffle the sum over instantons and anti-instantons to a summation over net topological number, $n_+ - n_-$, with
\begin{align}
\exp \ ( {-\mathcal{E} \, V \cdot T} )
&= \text{ const.} ~ \times
\nonumber \\
&
\hspace{-80pt}
\left( 1 + 2 
\sum\limits_{\substack{
n_+,  n_- \\
n_+ - n_- = 1 
}}
\frac{1}{n_+! ~ n_-!}
\left[ \int d^4 x_0 \int \frac{d \rho}{\rho^5} ~ C ~ \frac{1}{g^8}
\exp (-8\pi^2 / g^2(\rho)) \right]^{n_+ + n_-} + \ldots \right) &
\nonumber \\
&\simeq \exp \left( 2 \int d^4 x_0 \frac{d \rho}{\rho^5} C
\frac{1}{g^8} \exp (-8\pi^2 / g^2(\rho)) \right) \ . 
\label{eqn:EVT}
\end{align}
In the last line, we have used the identity
\begin{align}
\exp (2x) = \sum_{n_1} \sum_{n_2} \frac{1}{n_1!} ~ \frac{1}{n_2!} e^{n_1 x} `~ e^{n_2 x} \ .
\end{align}
From~\eqnref{EVT}, we note that $V \cdot T = \int d^4 x_0$ is the volume of Euclidean spacetime, and we thus conclude that the vacuum energy density from the instanton configurations is
\begin{align}
\mathcal{E} = -2 \int \frac{d\rho}{\rho^5} ~ C ~ \frac{1}{g^8}
\exp \left[ -8\pi^2 / g^2 (\rho) \right] \ .
\end{align}

Now, we consider the contribution to the vacuum energy density when the Yang-Mills Lagrangian (again in Minkowski metric) includes a $\Theta$ term,
\begin{align}
\mathcal{L} = -\frac{1}{4} G_{\mu \nu}^a G^{\mu \nu, a} + \Theta \frac{g_s^2}{32 \pi^2} G_{\mu \nu}^a \tilde{G}^{a, \ \mu \nu} \ .
\label{eqn:YangMillsTheta}
\end{align}
The only effect on the previous computation is a new factor of
$e^{i ~ \Theta ~ (n_+ - n_-)}$ in~\eqnref{dilute}, since the $G \tilde{G}$ term integrates to a net topological winding number.  Carrying this through, the sum over net winding number in~\eqnref{EVT} now evaluates to
\begin{align}
\exp \ ( {-\mathcal{E} ~ V \cdot T} )
&\simeq  \exp \left[ \left( e^{i ~\Theta}+ e^{-i ~\Theta} \right)
\int d^4 x_0 \int \frac{d\rho}{\rho^5} ~C~ \frac{1}{g^8}
\exp \left[ -8\pi^2 / g^2 (\rho) \right] \right] \ ,
\end{align}
and we obtain the vacuum energy density expression
\begin{align}
\mathcal{E} = -2 \cos \Theta \int \frac{d\rho}{\rho^5}
~C~ \frac{1}{g^8} \exp \left[ -8\pi^2 / g^2 (\rho) \right] \ .
\label{eqn:cosTheta}
\end{align}
Given the integral over instanton size is evaluated with an infrared cutoff, the vacuum energy density induced by instanton configurations in the dilute instanton gas approximation has a cosine dependence on $\Theta$.  We can now distill the PQ mechanism into simple energetics: by coupling the QCD axion to this non-perturbatively generated vacuum potential, the axion field acquires a vacuum expectation value that cancels the $\Theta$ parameter in~\eqnref{YangMillsTheta}, solving the strong CP problem.

\section{The strong CP problem in the Standard Model and the axion solution}
In the Standard Model, the non-observation of an electric dipole moment (EDM) for the neutron puts severe constraints on the $\bar{\Theta}$ parameter,
\begin{align}
    \bar{\Theta} \equiv \Theta + \arg \det Y_u Y_d \ ,
\label{eqn:barThetadef}
\end{align}
where $\Theta$ is the same parameter as in the Yang-Mills Lagrangian~\eqnref{YangMillsTheta} and $Y_u$ and $Y_d$ are the (unknown) original Yukawa matrices for up and down quarks in the Standard Model (SM).

The current constraint on the neutron EDM comes from the nEDM collaboration, with $|d_n| < 1.8 \times 10^{-26}$ $e$ cm~\cite{Abel:2020pzs}.  Following Ref.~\cite{Pospelov:1999mv}, this upper bound constrains the $\bar{\Theta}$ parameter since $d_n = C_{\text{EDM}} e \bar{\Theta}$, where $C_{\text{EDM}} = 2.4 \times 10^{-16}$ in the SM.  As a result, the current nEDM upper bound leads to \begin{align}
    \bar{\Theta} < 7.5 \times 10^{-11} \ ,
    \label{eqn:barThetabound}
\end{align} 
which quantifies the strong CP problem.  Returning to~\eqnref{barThetadef}, we emphasize that $\bar{\Theta}$ has two profoundly distinct origins.  First, the $\Theta$ parameter of QCD has a domain $(-\pi, \pi]$, and thus would generally be expected to be an $\mathcal{O}(1)$ number.  Second, the $Y_u$ and $Y_d$ matrices are similarly not measured in the Standard Model: we only know the eigenvalues of these matrices after rotating to the fermion mass basis, and we also know that the combination of the unitary matrices that diagonalize $Y_u$ and $Y_d$ gives the Cabibbo-Kobayashi-Maskawa matrix that has a Jarlskog invariant $CP$-violating measure of $3.08 \times 10^{-5}$~\cite{ParticleDataGroup:2022pth}.  Hence, the strong CP problem stems from the fact that two {\it unrelated} origins for $\bar{\Theta}$ must be radically aligned and canceled to achieve an acceptable value allowed by nEDM constraints, which is exacerbated by the large domain for each underlying phase parameter.  As an aside, we can mention that the massless up quark solution for the strong CP problem, albeit excluded by lattice measurements~\cite{Fodor:2016bgu}, is reflected in the structure of~\eqnref{barThetadef}.  Namely, if the up quark were massless, the $\arg \det Y_u Y_d$ term becomes {\it undefined}, since the argument of $0$ is arbitrary, reflecting an enhanced axial symmetry in the fermion mass basis that can be used to rotate $\Theta$.  As another aside, an alternative class of models solving the strong CP problem are Nelson-Barr models~\cite{Nelson:1983zb, Barr:1984qx}, where CP violation is introduced via spontaneous breaking at a high scale.  A recent review of Nelson-Barr models can be found in Ref.~\cite{Dine:2015jga}.  

In this primer, we focus of course on the axion solution to the strong CP problem.  As alluded to in the previous section, we can use the energetics of the $\bar{\Theta}$ parameter to cause the axion field to develop a tadpole and cancel the original $\bar{\Theta}$ as a result.  It is necessary and sufficient to couple the axion field to a colored fermion $\psi$ via an axial-vector current,
\begin{align}
\mathcal{L} = \left( \frac{\partial_\mu a}{f_a} \right) \bar{\psi}  \gamma^\mu \gamma^5 \psi \ ,
\label{eqn:axionABJ}
\end{align}
where we have introduced the decay constant $f_a$ on dimensional grounds.  We recognize that the classical shift symmetry for the axion is
preserved by this interaction term since the axion is derivatively coupled.  Crucially, the derivative coupling of the axion is precisely a current coupling to the Adler-Bell-Jackiw chiral anomaly for $\psi$~\cite{Adler:1969gk, Bell:1969ts}, and by using the equation of motion for $\psi$, we can 
realize the desired coupling of $a$ to the dual field strength tensor of the gluons,
\begin{align} 
\mathcal{L} = N \frac{a}{f_a} \frac{g_s^2}{32 \pi^2} G_{\mu \nu}
\tilde{G}_{\mu \nu} \ ,
\label{eqn:axionGGdual}
\end{align}
where $N$ denotes the PQ anomaly with respect to the color group.  

Comparing the $\bar{\Theta}$ term with~\eqnref{axionGGdual}, we remark that a finite shift in the $a$ field redefines the $\bar{\Theta}$ term, and hence $a$ acts as a {\it spurion} for $\bar{\Theta}$.  It is crucial to recall that since $a$ is a Goldstone field at the classical level, the lack of an origin in field space for $a$ is the necessary requirement for this spurion interpretation.  Using the spurion argument (see Refs.~\cite{Kim:2008hd, Kivel:2022emq}), we can now observe that below the scale of QCD confinement, instantons induce a periodic potential for the axion leading the axion field to acquire a tadpole offset that
minimizes the overall vacuum energy
and adjusts the $\Theta$ parameter to zero:
\begin{align}
\mathcal{L} = \Lambda_{\text{QCD}}^4 \cos \left(
\frac{N a}{f_a} + \Theta \right) \ ,
\end{align}
where $\Lambda_{\text{QCD}}$ is the scale of QCD confinement, about
200~MeV, and is formally related to the topological susceptibility of
QCD.  After shifting the axion field to absorb the constant $\Theta$
vacuum angle (and setting $N = 1$ to avoid domain wall problems), we perform a Taylor expansion of the cosine to obtain an
axion mass squared of $m_a^2 f_a^2 \sim \Lambda_{\text{QCD}}^4 \sim m_{\pi}^2
f_\pi^2$.


Since the axion decay constant $f_a$ is
generally required to be very large, $\mathcal{O} (10^{9})$ GeV from
supernovae constraints~\cite{ParticleDataGroup:2022pth}, the axion
mass is typically very small, with
\begin{align}
m_a = 5.70 (7) \left( \frac{10^9~\text{GeV} }{f_a} \right) \text{ meV}
\ .
\end{align}
Given there are no QCD states below this mass, the only possible
Standard Model final states allowed for axion decays are photons and
perhaps neutrinos.  The neutrino decay is typically discarded since
they may not be kinematically allowed and also the rate is suppressed
by the neutrino mass divided by $f_a$.  On the other hand, the
diphoton rate is dictated by the PQ anomaly with respect to the
electromagnetic field strength, 
\begin{align}
\mathcal{L} = \frac{a}{f_a} \left( \frac{E}{N} - 1.92 \right)
\frac{e^2}{16 \pi^2} F_{\mu \nu} \tilde{F}^{\mu \nu} \ ,
\end{align}
where we have rescaled the $f_a$ scale by $1/N$ to move the $N$
dependence purely into the electromagnetic coupling and the $-1.92$
value arises from the axion mixing with the QCD mesons.  In benchmark
axion models, the EM anomaly is fixed by the color and EM charges of
the fermions coupling to the axion, such as $8/3$ for the
Dine-Fischler-Srednicki-Zhitnitsky model~\cite{Dine:1981rt,
  Zhitnitsky:1980tq} and $E/N = 0$ for the
Kim-Shifman-Vainshtein-Zakharov model~\cite{Kim:1979if,
  Shifman:1979if}.  We remark that the lifetime of the axion exceeds
the age of the universe for $m_a < 20$~eV with $E/N =
0$~\cite{ParticleDataGroup:2022pth}, and thus light axions are a prime
cold dark matter candidate.

We close by noting that the Peccei-Quinn mechanism relies upon the classical nature of the Peccei-Quinn symmetry to a high degree in order for the axion to enjoy an instanton-induced potential that naturally aligns the axion tadpole to the desired $\bar{\Theta}$ cancellation.  But as a global $U(1)$ symmetry, the Peccei-Quinn symmetry is, at least, generally expected to be broken by higher dimensional operators suppressed by the Planck scale, notwithstanding possible explicit breaking terms arising from ultraviolet completions of the Standard Model that can include new sources of CP violation.  The difficulty in protecting the Peccei-Quinn symmetry from these possible ultraviolet corrections is called the axion quality problem.  An exemplary study of the axion quality problem and its requisite fine-tuned effects on axion phenomenology can be found in Ref.~\cite{Elahi:2023vhu}.  Furthermore, many contemporary studies focus on axion-like particles (ALPs), where the mass and decay constant relationship is not derived by the QCD topological susceptbility, and thus the ALP mass and its couplings to SM particles are taken as free parameters.  Primary references for the phenomenology of ALPs and their effective Lagrangian coupling to the SM include~\cite{Brivio:2017ije, Bauer:2017ris, Gavela:2019cmq}.

\section{Conclusions}

We have introduced the basic physics of axions as the angular mode
associated with a complex scalar field charged under a Peccei-Quinn
symmetry.  We have seen that the Peccei-Quinn anomaly with the color
gauge group leads to an instanton-induced potential that breaks the
classical, continuous shift symmetry into a periodic shift symmetry,
resulting in a light axion mass proportional to the QCD confinement
scale and inversely proportional to the axion decay constant.  The EM
anomaly and the mixing with QCD mesons dictate the axion couplings to
photons, leading to a rich set of experimental signals across decades
in couplings and masses.  Thus, since the diphoton coupling is
essentially fixed given the axion mass, axions serve as a
one-parameter model for dark matter.  We close by remarking that
axion-like particles relax the relationship between the mass and the
axion decay constant by using a non-QCD topological susceptibility to
determine their relationship, which helps to populate the entire
$(m_a, f_a)$ plane probed by axion experimental efforts.

\section*{Acknowledgments}
\label{sec:acknowledgments}

This work is supported by the Cluster of Excellence {\em Precision Physics, Fundamental Interactions and Structure of Matter\/} (PRISMA${}^+$ -- EXC~2118/1) within the German Excellence Strategy (project ID 39083149).
The author would like to thank the Fermilab theory group for its hospitality while this work was in completion.  The author would also like to thank the students of the Bad Honnef summer school on ``Ultralight Dark Matter" and the students of the Phenomenology Symposium 2023 for their helpful comments and questions.




\bibliographystyle{plain}

\end{document}